\documentclass{article}
\usepackage{ijcai18}
\usepackage{times}
\usepackage{xcolor}
\usepackage{soul}
\usepackage[utf8]{inputenc}
\usepackage[small]{caption}
\usepackage{latexsym}

\usepackage{graphicx}
\usepackage{amssymb,amsmath}
\usepackage{color}
\usepackage{fancyhdr}
\usepackage{nopageno}
\usepackage[ruled]{algorithm2e}
\usepackage{url}
\usepackage{hhline}
\usepackage{float}
\usepackage{enumitem}

\begin{document}


%

%
\title{Human Satisfaction as the Ultimate Goal in Ridesharing} 
\author{Chaya Levinger, Amos Azaria, Noam Hazon\\
	Department of Computer Science, Ariel University\\
	{chaya76@gmail.com,
	amos.azaria@ariel.ac.il,
	noamh@ariel.ac.il}
}
\maketitle

\begin{abstract}
Transportation services play a crucial part in the development of modern smart cities.  
In particular, on-demand ridesharing services, which group together passengers with similar itineraries, are already operating in several metropolitan areas.
These services can be of significant social and environmental benefit, by reducing travel costs, road congestion and co2 emissions. The deployment of autonomous cars in the near future will surely change the way people are traveling. It is even more promising for a ridesharing service, since it will be easier and cheaper for a company to handle a fleet of autonomous cars that can serve the demands of different passengers. 

We argue that user satisfaction should be the main objective when trying to find the best assignment of passengers to vehicles and the determination of their routes. Moreover, the model of user satisfaction should be rich enough to capture the traveling distance, cost, and other factors as well. We show that it is more important to capture a rich  model of human satisfaction than peruse an optimal performance. That is, we developed a practical algorithm for assigning passengers to vehicles, which outperforms assignment algorithms that are optimal, but use a simpler satisfaction model.

To the best of our knowledge, this is the first paper to exclusively concentrate on a rich and realistic function of user satisfaction as the objective, which is (arguably) the most important aspect to consider for achieving a widespread adaption of ridesharing services. 

\end{abstract}

\section{Introduction}
\label{sec:backg}

The National Household Travel Survey performed in the U.S. in 2009 \cite{santos2011summary} revealed that approximately 83.4\% of all trips in the U.S. were in a private vehicle (other options being public transportation, walking, etc.). The average vehicle occupancy
was only $1.67$ when compensating for the number of passengers (i.e., if two people travel in the same vehicale, their travel distance is multiplied by two). This extremely low average vehicle occupancy entails a very large number of vehicles on the road that collectively contribute to carbon dioxide emissions, fuel consumption, air pollution and an increase in traffic load, which in turn requires additional investment in enlarging the road infrastructure.
In recent years, ride hailing services such as Uber and Lyft have gained popularity and an increasing number of passengers use these services as one of their main means of transportation \cite{wallsten2015competitive}. 
Both Uber and Lyft are now also offering ride-sharing options, and other companies, such as Super-Shuttle and Via, are explicitly targeted at customers who want to share their ride.

The deployment of autonomous vehicles in the near future will have a significant impact on the way people are traveling. 
The implication of this revolutionary way of transportation is not fully known nowadays~\cite{guerra16}, but it is safe to claim that autonomous vehicles will have a positive effect on the development of ridesharing services. Indeed, it will be easier and cheaper for a company to handle a fleet of autonomous vehicles that can serve the demands of different passengers. It can also rule-out some negative human-driver factors, such as driver's fatigue from the long travels and the driver's inconvenience from having multiple pick-up and drop stops along his route. 

The basic challenge of a ridesharing service is how to assign the passengers' requests for a ride to vehicles and define the routes for the fleet of vehicles in an optimal manner. This problem belongs to the generic class of Vehicle Routing and scheduling Problems (VRPs), which have been extensively studied over the past $50$ years, mainly in the operation research and transportation science communities. Several variants with different characteristics have been developed. For example, the initial formulation of the VRP assumes that the environment is \textit{static}, i.e., all requests are known before-hand and do not change thereafter~\cite{dantzig1959truck}. The more complex variants, including the rideharing problem, are \textit{dynamic}, where real-time requests are gradually revealed along the service operating time. In this setting the assignment of passengers to vehicles and the determination of vehicles' routes may be adjusted when they are already in transit~\cite{psaraftis2016dynamic,shen2016dynamic}. Arguably, a major criterion that characterizes each variant of the VRP is the objective function. It is very common to consider objectives from the service provider's perspective, for example, minimizing the total distance travelled~\cite{secomandi2000comparing}, minimizing the fleet size~\cite{diana2006model,secomandi2009reoptimization}, or maximizing the service provider's profit~\cite{campbell2005decision,parragh2014dial}. However, as noted by Cordeau and Laporte~\cite{cordeau2003tabu}, one should be interested not only in
minimizing the operating costs for the service provider but also in maximizing the quality of the service and the user satisfaction.

Many works integrate quality of service and user satisfaction considerations as additional constraints of the problem. For example, a time window restricts the waiting time a passenger is willing to face before being picked up~\cite{jaw1986heuristic,el2010vehicle}, and it is usually combined with a bound on the maximum user ride time~\cite{paquette2009quality}.
In addition, there are several works that combine the aforementioned operational objectives with the objective of maximizing the user satisfaction (or its antonym, minimizing the user inconvenience). The  common interpretation for user satisfaction is the minimization of the total user on-board (ride) time and the total user waiting time~\cite{psaraftis1980dynamic}, the extra riding time due to ride-sharing~\cite{lin2012research}, or the amount of deviations from desired departure and arrival times~\cite{fu1999line,yang2013multi}. 
However, to the best of our knowledge, there are no works that exclusively focus on maximizing a complex user satisfaction function, which captures the traveling distance, cost, and other factors as well.

We investigate a comprehensive human-centric approach for the ridesharing problem. 
Our basic claim is that the user satisfaction should be the main objective of the ridesharing service. Moreover, the model of user satisfaction should be rich enough to capture the complex interdependencies among several factors. Therefore, we develop a method for maximizing a complex user satisfaction function.

One approach for handling a rich objective function is to treat its factors as multiple objectives. Indeed, there are several methods in the literature on VRP for handling multiple objectives. The most common approach is to aggregate the objectives into a single weighted-sum objective function~\cite{molenbruch2017}, and advanced utility model may be used for modeling the interactions between the objectives~\cite{lehuede2014multi}. Additional strategies define hierarchical objective function~\cite{schilde2011metaheuristics}, or return a set of non-comparable solutions which do not weakly dominate each other~\cite{parragh2009heuristic,molenbruch2017multi}.
Since our rich objective function models user satisfaction, we propose a different, human-centric, approach.
Specifically, we investigate machine learning methods for modeling the rich satisfaction function from real humans.

Modeling human behaviour is not a easy task, and a theoretical model might fail to accurately capture real human behaviour. We therefore ran experiments with actual humans and build a deep learning based function to estimate user satisfaction. We introduce Simsat, an algorithm for assigning passengers to vehicles while maximizing a complex user satisfaction function as the objective, for the ridesharing last mile variant \cite{cheng2014mechanisms} setting. We show that Simsat outperforms optimal assignment methods that use a simpler objective function, indicating that it is more important to obtain a richer model of user satisfaction, than improving the performance of the assignment algorithm.




\section{Related Work}


We will briefly review the current literature on the broad class of Vehicle Routing and scheduling Problems (VRPs), to place our ridesharing problem in an appropriate context.  
The VRP was first introduced by~\cite{dantzig1959truck}.
The growing body of research on routing problem over the past $50$ years has led to the development of several research communities, which sometimes denote the same problem types by various names. In particular, the traditional VRP and some of its extensions deal with finding an optimal set of routes for a fleet of vehicles to traverse in order deliver or pickup some goods to a given set of costumers. We refer to the comprehensive survey of~\cite{parragh2008a} on this class of problems, which they denote by Vehicle Routing Problems with Backhauls (VRPB). A more recent survey, that also defines a taxonomy to classify the various variants of VRP by $11$ criteria, is given by~\cite{psaraftis2016dynamic}. A second class of problems, that is denoted by Parragh et al.\ as Vehicle Routing
Problems with Pickups and Deliveries (VRPPD), deal with all those problems where goods are transported between pickup and delivery customers. We refer to the survey of~\cite{parragh2008b} on this class of problems. One subclass of VRPPD compromises the dial-a-ride problem (DARP), where the goods that are transported are passengers with associated pickup and delivery points.
As noted by~\cite{cordeau2003tabu}, the DARP is distinguished from other problems in vehicle routing since transportation cost and user inconvenience must be weighed against each other in order to provide an appropriate solution. Therefore, the DARP typically includes more quality constraints that aim at capturing the user's inconvenience. We refer to a recent survey on DARP by~\cite{molenbruch2017}, which also makes this distinction. 


A domain closely related to ride-sharing is car-pooling. In this domain, ordinary drivers, may opt to take an additional passenger on their way to a shared destination. The common setting of car-pooling is within a long-term commitment between people to travel together for a particular purpose, where ridesharing is focused on single, non-recurring trips. Indeed, several works investigated car-pooling that can be established on a short-notice, and they refer to this problem as ridesharing~\cite{agatz2012optimization}. We stress that in our ridesharing problem, similar to the DARP setting, there is a central organization that owns the vehicles, and they thus do not have  their own travel plans. 

%

\section{Basic Model and Notation}


Informally, the ridesharing problem consists of a weighted graph, requests given by passengers, each with an origin and a destination that are both nodes in the graph, and a set of vehicles, each with a given capacity. All the vehicles are assumed to be operated by a central entity. In this paper we focus on the last mile variant \cite{cheng2014mechanisms} setting. In this variant it is assumed that all the passengers are positioned at the same origin location (e.g. airport), where all the vehicles are also located, and must be taken to their final destination.
The problem requires assigning travel routes (on the graph) to vehicles, in order to satisfy these passenger requests while optimizing a given objective function. In our work, we concentrate on the objective of maximizing the user satisfaction function.

Let $n$ be the number of passengers (and thus the number of requests).
We assume that the service provider incurs a fixed cost per minute of travel, $M$, that encapsulates the full operation cost including any desired revenue. 
For example, if the fuel, tolls and any maintenance costs are estimated at $x$ dollars per minute of travel, and the service provider commits to receiving only a certain percentage of the user payment (as its revenue), $\rho$, $M$ equals $\frac{x}{1-\rho}$.
The service provider is free to determine how to distribute this cost among the passengers, and in section \ref{sec:paymentFun} we discuss the properties of this payment function.

Every user, $u\in U$, is assumed to have a primary travel time $t_o(u)$ and distance $d_o(u)$, which are the time and distance, respectively, it would take the user to reach her destination had she received a direct ride. Consequently, we define for each user a primary travel cost $c_o(u)=M\cdot t_o(u)$.
We will also add a fixed constant to $t_o(u)$ that represents waiting time of the passenger had she received a direct ride. Given an assignment, $P$, and a user $u$, the actual travel time of the user is denoted by $t_P(u)$, the actual travel distance is denoted by $d_P(u)$, the actual amount paid by the user is denoted by $c_P(u)$, and the user satisfaction is denoted by $s_P(u)$.

\subsection{The Human Satisfaction Function}
Our definition of the objective of the ridesharing problem is to find an assignment $P$ that maximizes the sum on all user satisfaction, i.e., $\sum_{u\in U}s_P(u)$ (or, equivalently, the average satisfaction $\frac{1}{n}\sum_{u\in U}s_P(u)$). For simplicity, in the last mile problem we assume that a passenger traveling alone has some baseline satisfaction level (``neither satisfied nor disatisfied'') from the trip.
%
%
%
%
%
Satisfaction factors may include any or all of the following:
\begin{itemize}
\setlength{\itemsep}{2pt}
\item {\bf Travel cost}. $c_P(u)$.
\item {\bf Actual travel time}. $t_P(u)$. The travel time may also depend on other parameters, for example, a user may care more about travel time during the weekend.
\item {\bf Extra travel time}. $t_P(u)-t_o(u)$.
\item {\bf Actual travel distance}. $d_P(u)$.
\item {\bf Extra travel distance}. $d_P(u)-d_o(u)$ 
\item {\bf Total number of passengers}. Users may rather travel alone. The more passengers a ride may have the more inconvenient it may become to each of the passengers.
\item {\bf User's seat}. In every vehicle with a constant capacity some seats may be preferred over others. For example, if a vehicle with $5$ seats reaches its full capacity, most people prefer siting on the front seat rather than the middle back seat.
\item {\bf Working status / Occupation}. Unemployed passengers may be willing to travel longer in exchange for a lower cost.
\item {\bf User demographic information}. Depending on the user demographic group (e.g. age, gender, annual income), users may care more about the other factors. For example, young people may not mind traveling longer if they save a few dollars, but people in their 40's may be more concerned about their time.  Some of this information may be extracted from an image of the user. Note that this factor allows to define a more personalized function, since different users might end up with different satisfaction values for identical rides.
\end{itemize}



\subsection{Payment Function for the Last Mile Problem}
\label{sec:paymentFun}
Given $n$ passengers and their destinations, who travel in a single vechile, an assignment $P$ determining the drop-off path (i.e. the order of which the passengers are dropped-off), the time it takes to reach each destination under this assignment ($t_P(u)$), the total cost of the ride-shared trip, and the cost ($c_o(u)$) and travel time ($t_o(u)$) that each of the passengers would have encountered had they traveled alone, the payment function determines how much each passenger must pay ($C_P(u)$) when all passengers share the ride.
We define the following axioms on the payment function:
\begin{enumerate}
    \item The aggregated payment from all passengers should exactly cover the cost of the trip.
    \item $n$ passengers traveling to the same destination split the trip cost equally.
    \item Given two passengers with the same distance from the source, the passenger who is dropped off second should pay less than the passenger who was dropped off first.
    \item Given a passenger that is "on the way" to another passenger, both passengers should pay strictly less than what they would pay had they traveled alone (even though, both passengers do not travel any longer).
\end{enumerate}

The following payment schedule satisfies all the axioms above:
Let $\alpha, \beta \in R^+$, we define the user's gain (given an assignment $P$) as:
\begin{align}
\label{sat:simple}
g_P(u) = (\alpha t_o(u) + \beta c_o(u)) - (\alpha t_P(u) + \beta c_P(u))
\end{align}
If we define the inconvenience of a user, $i_P(u)$, as $\alpha t_P(u) + \beta c_P(u)$ (and similarly, $i_o(u)$ is $\alpha t_o(u) + \beta c_o(u)$), Equation \ref{sat:simple} can be simplified to $g_P(u) = i_o(u) - i_P(u)$. The payment schedule sets $c_P(u)$ such that all passengers traveling in the same ride have the same $g_P(u)$, and the sum of $c_P(u)$ equals the cost of the whole ride. Equation \ref{sat:simple} implies a simplified view of human behavior with the single concept of ``time is money'', and it further assumes a linear relation between the two.

\section{Satisfaction Model Learned from Humans}
\label{sec:satFromMTurk}
In order to develop a more realistic human satisfaction model, we use machine learning techniques based upon data collected from humans. To this end, we solicited $414$ human subjects from Mechanical Turk to obtain satisfaction level data. Based on this data, we use deep learning to build a satisfaction model.

\subsection{Data Collection}
The subjects were first asked to provide the following personal details: year of birth, gender and whether they were employed or unemployed.

Our satisfaction model tries to predict the relative satisfaction, that is, how much a passenger traveling by shared-ride is more or less satisfied than the same passenger traveling in a private ride. However, asking users to provide their relative satisfaction is unrealistic, and we thus split every travel scenario into two sub-parts. In the first part we asked the subjects to determine their satisfaction level from a direct private ride to some destination. In the second part the subjects were asked to determine their satisfaction level from a shared ride to the same destination. 
Specifically, in the first part of each scenario we described a direct private ride with a given time (random number between 5 minutes and one hour) and price (dollar per minute).
In the second part of each scenario we described a shared ride to the same destination, where we varied the travel time and cost.
Travel time of a shred ride can never be shorter than a direct private ride, and we thus uniformly sampled a number from $\{ 1, 1.1, 1.2, 1.3, 1.4, 1.5, 1.75, 2, 3,  4 \}$ and multiplied it by the direct private travel time.
The cost of a shared ride should be lower than the cost of a direct private ride. In the optimal sharing scenario, assuming a 4 passenger vehicle (excluding the driver), there could be 4 passengers traveling to the same destination;  in this case the cost will be reduced by a factor of 4. We thus divided the direct private ride's cost by a number uniformly sampled from $\{1.1, 1.2, 1.3, 1.4, 1.5, 1.75, 2, 3,  4\}$.
In addition, we randomly sampled the number of additional passengers from $\{1,2,3\}$, and we randomly sampled the user's seat from $\{$front passenger, middle back, right back, left back$\}$.
The subjects could choose one of seven satisfaction levels on a Likert scale, between very satisfied (7) to very dissatisfied (1) with the middle being `neither satisfied nor dissatisfied' (4).

Each subject was asked a total of six travel scenarios (each with a private and a shared ride). 
In order to eliminate subjects that may be selecting satisfaction levels at random, we added two sanity check scenarios. In these scenarios the cost of the shared ride was \emph{more} expensive than the private ride. Since it is unreasonable for a person to be more satisfied with a shared ride, being both longer and more expensive, we have disqualified any subject who expressed her satisfaction in this question to be higher than her satisfaction from the private ride of that scenario. $131$ subjects failed one of these sanity tests, and were removed from our analysis.

$26$ subjects refused to answer the personal questions and were eliminated from our analysis as well. 
Of the remaining $257$ subjects $147$ were females and $110$ were males. Their age ranged from $19$ to $67$ with an average of $32.3$ and a median of $31$. $195$ were employed while $62$ were unemployed. Each of the $257$ subjects had $4$ scenarios, resulting with $1028$ data-points.
The average satisfaction for a private trip was $5.01$ and a shared ride was $4.41$. 

\subsection{Deep-Learning Based Model}
Using the collected data we consider deep learning based models with a varying depth to find a good satisfaction model, that is, a model that will accurately predict user satisfaction levels of a new user, based on different features of the user and the user trip. We split the data into train, validation (dev) and test sets. We use mean squared error to measure the performance of each model. The neural network depth varied from 1 (linear regression) to 4 (3 hidden layers). Each hidden layer consisted of $100$ neurons. We used early stopping \cite{prechelt1998automatic}, i.e., we used the validation set to determine when to stop training. Table \ref{tbl:sat_model} presents the results obtained by each of these models. 
Since the 2-hidden layers model performed best, we use it as our satisfaction function, this model achieved a root mean squared error of $1.75$ on the test set.
We set the satisfaction of a private ride to `neither satisfied nor dissatisfied' (4).

\begin{table}
 \begin{tabular}{|c | c c|} 
 \hline
 Model & Train-error & Validation error \\  
\hhline{|=|==|}
 Linear regression & 1.680 & 1.734 \\ 
 \hline
 1 hidden layer & 1.621 & 1.711 \\
 \hline
 2 hidden layers & 1.592 & {\bf 1.696} \\
 \hline
 3 hidden layers & 1.480 & 1.785 \\ 
 \hline
\end{tabular}
\caption{Train and validation error for the different model depths. Values indicate the root mean squared error (RMSE) of each model.}
\label{tbl:sat_model}
\end{table}






\section{Experimental Evaluation}
\label{sec:preliminary}
In this section we present a stochastic algorithm for the last mile variant. 
We assume that there are a sufficiently large number of vehicles so that any request could be satisfied, and that the capacity of each vehicle is $4$ passengers.
We compare the performance of our algorithm, in terms of user satisfaction, to an optimal algorithm that uses a simpler satisfaction function. We show that the  algorithm outperforms the optimal algorithm, and that emphasizes the importance of capturing a rich model of human satisfaction.

\subsection{Stochastic-Based Satisfaction Algorithm (Simsat)}
We now present, a practical algorithm for assigning passengers to vehicles with the objective of maximizing the sum on user satisfaction: Simsat (Algorithm \ref{alg:simsat}). Simsat runs Floyd-Warshell on the graph at the initialization, to obtain the minimal travel time between every two vertices. Simsat then runs its main procedure as follows. Simsat shuffles all passengers and assigns every passenger to the vehicle that maximizes the current satisfaction sum. For computing the total satisfaction of a single vehicle (the SatFunc function in Algorithm \ref{alg:simsat}), we use the nearest neighbor algorithm (the greedy approach) for ordering the passengers drop-offs (based upon the Floyd-Warsell matrix). 
The main procedure is repeated multiple times ($n^2$) and the assignment that yields the maximal total satisfaction is selected. The complexity of Simsat is clearly $O(n^4)$. The number of times the main procedure is repeated can vary; the more times it is repeated the higher the expected performance. Therefore, Simsat is an any-time algorithm.
\begin{algorithm}[hbpt]
\caption{The Simsat algorithm}
\label{alg:simsat}
\SetAlgoLined
\textbf{Input}:
A graph (Graph), with source vertex\\
Passenger destinations list (Passengers), \\
A satisfaction function that returns the \emph{total} satisfaction of \emph{all} passengers in a vehicle (SatFunc),\\
 \KwResult{An assignment of all passengers to vehicles.} 
 Compute Floyd-Warshell on Graph\;
 MaxSum := 0\;
 \For {i:=0 to $n^2$}{
 Shuffle Passengers\;
 SatSum := 0\;
 Clear CabList\;
 \For{CurrentPassenger : Passengers}{
 MAX := -1\;
 \For{CurrentCab : CabList}{
 \If{CurrentCab not full}
 {CurrentSat := SatFunc(CurrentCab)\;
 Add CurrentPassenger to CurrentCab\;
 SatWithCurPass = SatFunc(CurrentCab)\;
 Remove CurrentPassenger from CurrentCab\;
 \If{(SatWithCurPass - CurrentSat) larger than MAX }
 {OptimalCab := CurrentCab\;
 MAX := SatWithCurPass - CurrentSat}
 }
 }
 \eIf {MAX $<$ 0}
 {
 Add CurrentPassenger to newCab\;
 Add newCab to CabList\;
 }
 {
 Add CurrentPassenger to OptimalCab\;
 SatSum += MAX\;}
 }
  \If{(SatSum $>$ MaxSum)}
 {MaxSum := SatSum\;
 OptimalFullAssignment := CabList\;}
 }
\end{algorithm}
\subsection{Optimal Algorithm}
We use the following method to obtain the optimal assignment. First, the algorithm runs Floyd-Warshell on the graph. The algorithm then solves a coin-change problem (\cite[p.~171]{harris2008combinatorics}) to obtain all possible ways to split the number of passengers into vehicles. For example, when $n=10$, we get $\{3,3,3,1\}$, $\{2,2,2,2,2\}$, $\{4,4,2\}$ etc. For each splitting option the algorithm iterates over all possible assignments (we explicitly handle multiple vehicles with the same number of passengers, since it does not matter which group of passengers travels in which vehicle if they are in the same size). For example, for a group of $\{4,3,3,2,2,2\}$, it first iterates over all assignments of 4 passengers (there are $\binom{n}{4}$ such assignments), then, recursively calls the assignment function with $\{3,3,2,2,2\}$ and the remaining passengers. The recursive call iterates over all possible assignments of three people to two vehicles and preforms a recursive call with the remaining vehicles and passengers.
For each vehicle, the algorithm computes all possible options for dropping off its passengers (this is done once for each set of users), and, based upon the Floyd-Warshell matrix and the satisfaction model, selects the most efficient travel order. 

\subsection{Data}
\begin{figure}
\centering
\includegraphics[width=8cm]{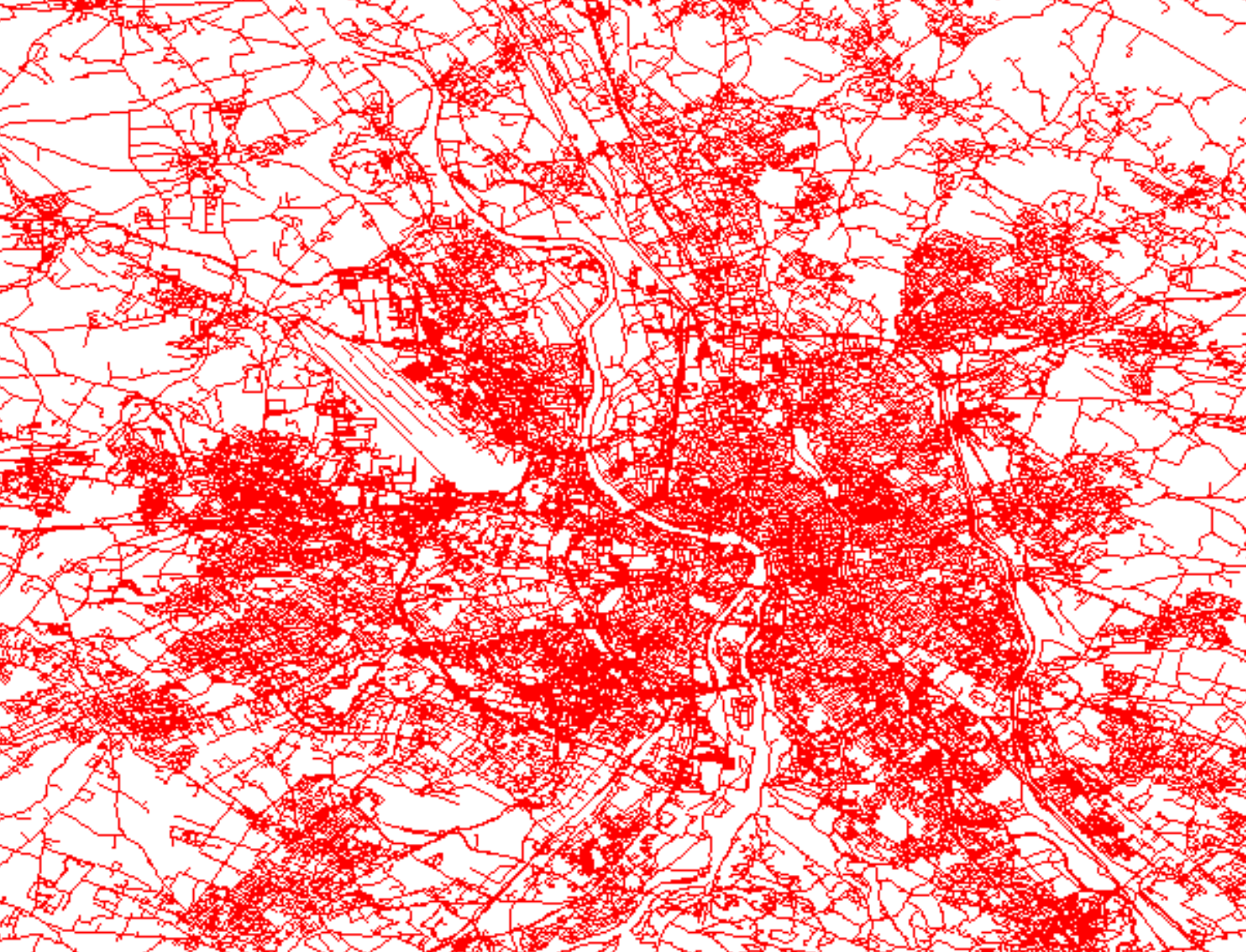} \caption{A graph created from a map of the city of Toulouse, France.}
\label{fig:ToulouseGraph}
\end{figure}

We considered two different types of graphs, a randomly generated graph and a more realistic graph, the city of Toulouse, France\footnote{obtained from \url{https://www.geofabrik.de/data/shapefiles_toulouse.zip}}.
The random graph was created by placing $35$ vertices uniformly on the plane. We then randomly chose a pair of vertices, and connected them with an edge with a probability that is proportional to their distance. The weight of each edge was determined by the air-distance multiplied by a random number (uniformly sampled) between 1 and 2, to model topological variance.
The graph of the city of Toulouse is presented in Figure \ref{fig:ToulouseGraph}. This graph includes the actual distances between the different vertices. We cropped the graph to $40,000$ vertices, by running Dijkstra algorithm starting at the airport, sorting all vertices by their distance from the airport, and removing all farther away vertices (including those that are unreachable).

\subsection{Experimental Settings}
Being a last mile problem, we set the origin vertex to be the same for all passengers. That is, 
in the random graph we randomly generated an origin vertex. In the city of Toulouse the graph includes the Toulouse-Blagnac airport, and it was used as the origin vertex. The destination vertices were randomly sampled for every passenger using a uniform distribution over all vertices.

In the payment schedule we set $\alpha$ to $0.3$ and $\beta$ to $1$ for the gain function. $\alpha$ and $\beta$ were set according to the average U.S. wage. That is, average annual hours worked per worker in U.S. at 2016 was 1783, and the average annual income in the U.S. per worker in 2016 was \$31,099. Dividing the two we get \$17.5 per hour, or approximately \$0.3 per minute. We set the average speed to 60 kph, and the cost per $km$ travel distance to $\$1 $.
We tested 5 assignment algorithms:
\begin{enumerate}
    \item The optimal algorithm with the full satisfaction function (developed in section \ref{sec:satFromMTurk}).
    \item Simsat with the full satisfaction function.
    \item The optimal algorithm with simpler satisfaction functions:
    \begin{enumerate}
        \item Travel cost only.
        \item Travel time only.
        \item Time and cost according to the payment function (that is, the gain function is used as a substitute for user satisfaction).
    \end{enumerate}
\end{enumerate}
All the algorithms were evaluated with the complete satisfaction function developed in section \ref{sec:satFromMTurk}, regardless of the function actually used by the assignment algorithm.

\begin{figure}
\centering
\includegraphics[width=8.5cm]{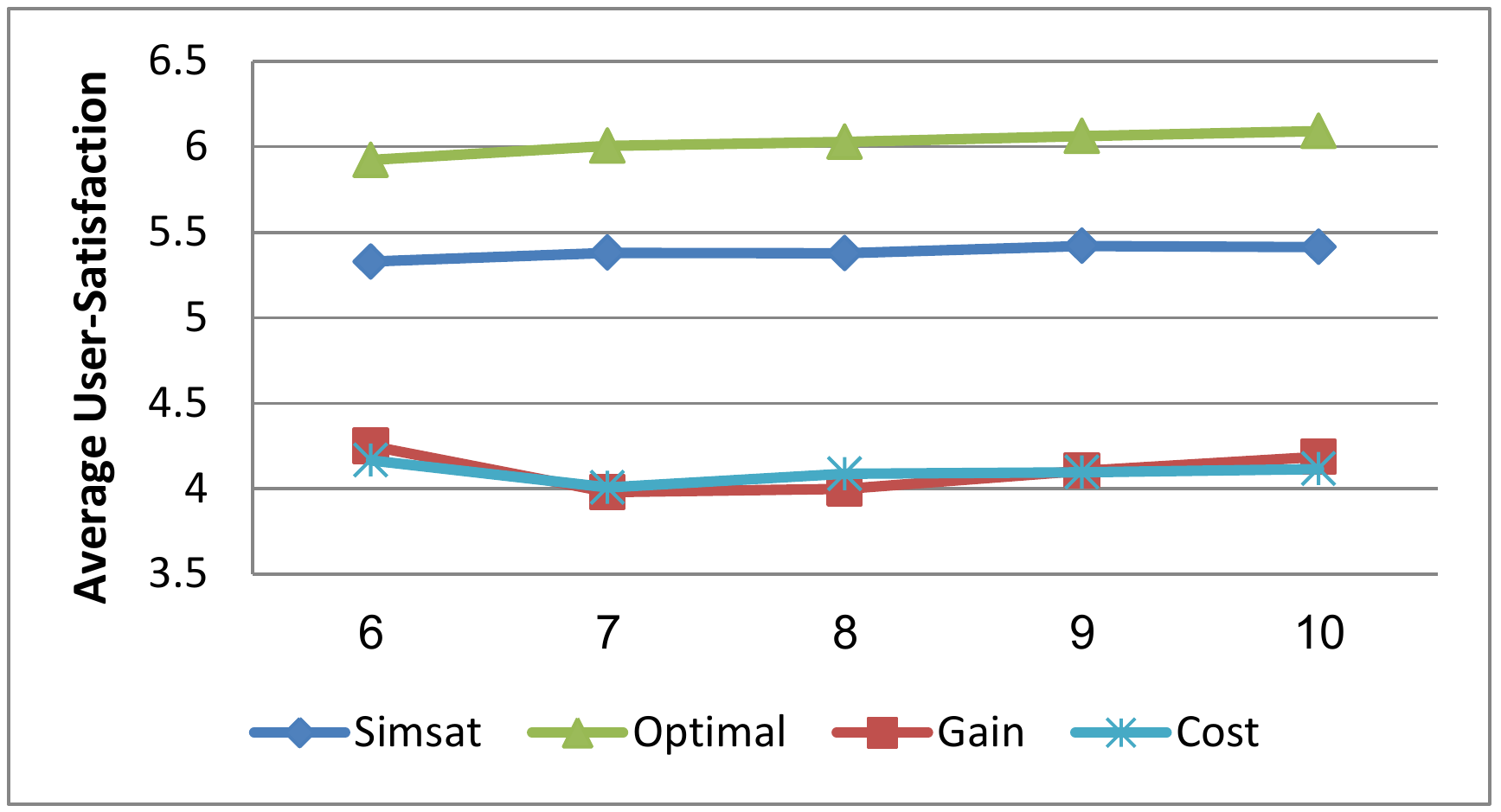} \caption{Average satisfaction for 6, 7, 8, 9 and 10 passengers when using a random graph.}
\label{fig:RandomGraphRes}
\end{figure}

\begin{figure}
\centering
\includegraphics[width=8.5cm]{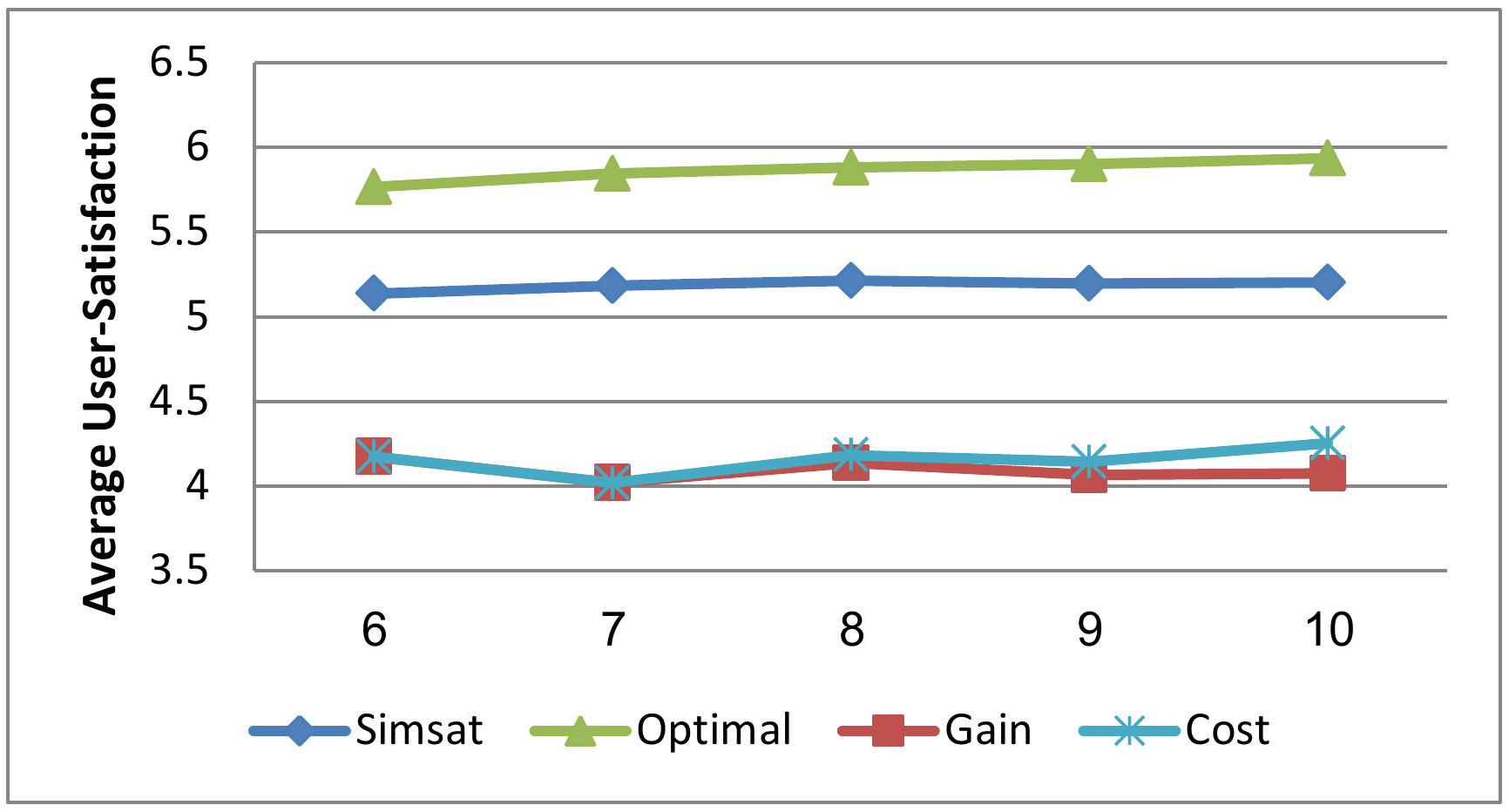} \caption{Average satisfaction for 6, 7, 8, 9 and 10 passengers when using a map of the city of Toulouse, France.}
\label{fig:ToulouseRes}
\end{figure}

\subsection{Results}
Figure \ref{fig:RandomGraphRes} presents the average satisfaction for 6,7,8,9 and 10 passengers when using a random graph, and Figure \ref{fig:ToulouseRes} presents the results for the city of Toulouse. The results were obtained by averaging over $100$ samples of passenger destinations. The results for the optimal method using travel time only were omitted, as it constantly yields an average user satisfaction of $4$ (since it assigns a private vehicle to each and every passenger).
As depicted in both figures, our satisfaction oriented assignment method (Simsat) obtains results that are quite close to the optimal assignment. Simsat's average satisfaction level is much closer to the optimal assignment than that of the optimal assignments using a simpler user-satisfaction model. These results indicate that it is more important to obtain a richer model of user satisfaction, than improving the performance of the assignment algorithm. That being said, we do not disregard the importance of improving the performance of the assignment algorithm and do intend to pursue additional algorithms that may perform better.

\section{Conclusions \& Future Work}

Ridesharing has a true potential for improving the quality of life for many people~\cite{cici2013quantifying}, and it is part of the general concept of sharing economy that is being evolved nowadays. However, despite both Uber and Lyft offering ride-sharing options, not many users elect to share their rides with additional passengers~\cite{motherboard2016,RSGsurvey2017}. 
Following the statement by Carnegie~\cite[p.~37]{carnegie1936win}, "There is only one way to get anybody to do anything. And that is by making the other person want to do it.", we believe that the key ingredient required for a widespread adaptation of ridesharing is to focus on user satisfaction. 

The importance of the paper lies in its being the first to exclusively concentrate on a rich and realistic function of user satisfaction as the objective, which is (arguably) the most important aspect to consider for achieving a widespread adaption of ridesharing services. We use deep learning to model user satisfaction based upon data collected from actual human subjects. We present a satisfaction oriented assignment method (Simsat), and show that it outperforms optimal assignments using a simpler user-satisfaction model. These results indicate that it is more important to obtain a richer model of user satisfaction, than improving the performance of the assignment algorithm. 

In future work we intend to extend our model to the more general ridesharing schenario, where people may have different origins. We also intend to build a game that will simulate an actual ride for the subjects; this should allow us to obtain more exact satisfaction levels. This game could include additional travel information such as the other passengers in the trip, and allow the subject to select her seat when entering a vehicle. Since users will be playing the game more than once, the satisfaction model can be further improved by personalization, taking into account user's feedback on previous rounds.

{\small

\bibliography{researchProp}
\bibliographystyle{named}

}

\end{document}